\documentclass[aps,preprint,showpacs,floatfix,prl]{revtex4}
\usepackage{bbding}
\usepackage{epsfig}
\usepackage{graphicx}
\usepackage{dcolumn}
\usepackage{bm}
\usepackage{ulem}
\usepackage{amssymb}
\usepackage{amsmath}   
 
\begin{document}
\title{ Harnessing synthetic gauge fields for maximally entangled state generation }
\author{S. A. Reyes$^1$, L. Morales-Molina$^1$, M. Orszag$^1$ and D. Spehner$^{2,3}$}
\affiliation{$^1$Instituto de F\'isica, Pontificia Universidad Cat\'olica de Chile, Casilla 306, Santiago 22, Chile}
\affiliation{$^2$Universit\'e Grenoble 1 and CNRS, Institut Fourier UMR5582, B.P. 74, 38402 Saint Martin d'H\`eres, France}
\affiliation{$^3$Universit\'e Grenoble 1 and CNRS, Laboratoire de Physique et Mod\'elisation des Milieux Condens\'es UMR5493, B.P. 166, 38042 Grenoble, France}

\begin{abstract}
We study the generation of entanglement between two species of neutral cold atoms living on an optical ring lattice, where each group of particles can be described by a $d$-dimensional Hilbert space (qu$d$it). Synthetic magnetic fields are exploited to create an entangled state between the pair of qu$d$its. Maximally entangled eigenstates are found for well defined values of the Aharonov-Bohm phase, which are zero energy eigenstates of {\it both} the kinetic and interacting parts of the Bose-Hubbard Hamiltonian, making them quite exceptional and robust against certain non-perturbative fluctuations of the Hamiltonian. We propose a protocol to reach the maximally entangled state (MES) by starting from an initially prepared ground state. Also, an indirect method to detect the MES by measuring the current of the particles is proposed.
\end{abstract}

\maketitle

{\it Introduction}.---
Entanglement  is one of the most important and unique features of quantum mechanics leading to many applications such as teleportation and quantum cryptography \cite{four}. It may also be exploited for quantum metrology applications such as enhancing the precision of atomic interferometers used for time measurements and ultra-small signal detection \cite{interference}. Entanglement is a key ingredient in quantum information processing where the fundamental units are the so called quantum bits (qubits), corresponding to simple two-level systems. Generalizations of this concept to systems with $d$ dimensional Hilbert spaces (qu$d$its) have been the subject of great interest in recent years. For instance, it has been demonstrated that maximally entangled states among a pair of qudits violate local realism theories stronger than qubits \cite{PRL_Zeilinger}. Entanglement is a delicate quantum feature so its generation, protection, propagation, and distribution have been the subject of 
intense research over the last decades \cite{measure}.  In particular, it has been pointed out \cite{three} that entangled states are much more fragile than uncorrelated states, making difficult their use in applications.
Consequently, it remains a major challenge to generate robust entangled states since engineering experimental setups to observe quantum correlations requires high coherence and a precise control on the Hamiltonian. 

Recent experimental advances in the manipulation of ultracold atoms in optical lattices make them one of the most promising candidates due to the controllability of the system dimension \cite{Ober} and precise tunability of the interaction among particles \cite{FR}. 
In particular, ultracold atom systems have proven to be ideal devices for high precision interferometry \cite{Ober0}. For instance, entangling atoms in an optical lattice can reduce the noise in a clock measurement \cite{one}. 
In addition, current developments have demonstrated that Raman-assisted tunneling can be used to implement synthetic magnetic fields for neutral atoms in optical lattices \cite{NJP_Jak,RMP_gauge,PRL_2011_Bloch,NAT_Lin,PRL_Gold_1,PRL_Gold_2}. The atoms in the lattice can therefore acummulate a non-adiabatic Berry phase as they hop from site to site, usually referred to as the Peierls phase \cite{PRL_Spiel}. Similarly, such a phase can be engineered by applying a suitable periodic force \cite{AC_force}. These experimental tools open new routes for the observation of striking physics such as the quantum Hall effect \cite{PRL_2013-Ketterle,PRL_2013_Bloch} and novel topological phenomena \cite{PRL_Gold_2}.


In the present letter we propose a novel approach to produce maximally entangled states (MES) between two species of neutral atoms in an optical ring lattice by using synthetic magnetic fields, where each species has a fix atom number and constitutes a qu$d$it. We develop a protocol by which the MES can be reached through an adequate time variation of the Peierls phase. Furthermore, we show that the MES produced in this way is robust against certain fluctuations in the Hamiltonian parameters. We propose a way to measure the amount of entanglement between the atoms of the two species by looking at the particle current in the ring.

{\it The model}.---
Consider two species of interacting condensed atoms moving along a ring. This may be accomplished either by using the same isotope with different internal states \cite{Ober2} or by using two kinds of atoms, such as $^{87}$Rb and $^{40}$K \cite{Fesch}. In its simplest form, the dynamics of the system is governed by a Bose-Hubbard Hamiltonian, 
\begin{equation} \label{Eq:Hamiltonian}
\hat{H}=\hat{K}_A + \hat{K}_B + \hat{H}_{int}.
\end{equation}
As mentioned before, the inclusion of a synthetic magnetic field results in an acquired quantum mechanical phase by the atoms as they hop from site to site. Then, the kinetic part for each species $D=A,B$ is written as  
\begin{equation}\label{Eq:Binding}
\hat{K}_D=-C\sum_{j=1}^{L} (e^{i\phi_D} \hat{d}_{j}^{\dagger}\hat{d}_{j+1}^{\phantom{\dagger}}+ e^{-i\phi_D} \hat{d}_{j+1}^{\dagger}\hat{d}_j^{\phantom{\dagger}}),
\end{equation}
where $C$ is the tunneling strength, $\phi_D$ is the Peierls phase for species $D$, $L$ is the number of sites and $\hat{d}_{j}^{\dagger}$ ($\hat{d}=\hat{a},\hat{b}$) is the particle creation operator at site $j$. Since the particles live on a ring they obey periodic boundary condition $j+L=j$, so that $d_{L+1}^\dagger = d_1^\dagger$. The interaction part of the Hamiltonian is given by 
\begin{equation}\label{Eq:HamSpecie}
H_{int}= \frac{U}{2} \sum_{j=1}^{L} (\hat{n}^2_{Aj}+\hat{n}^2_{Bj}) - V \sum_{j=1}^{L}  \hat{n}_{Aj}\hat{n}_{Bj},
\end{equation}
where $\hat{n}_{Dj}=d_j^{\dagger}d_{j}$ is the particle number operator for species $D$. Interactions can be accurately tuned experimentally via Feschbach resonance \cite{Fesch,FR}. 
In this work we consider repulsive on-site intra-species interactions with the same magnitude $U>0$ for both species and attractive on-site inter-species interactions with magnitude $V>0$. 

Consider now a fixed number $N$ of particles for each species, such that both Hilbert spaces for the atoms $A$ and $B$ have dimension
\begin{equation} \nonumber
 d =  \left( \begin{array}{c}
N+L-1 \\
N \end{array} \right).
\end{equation}
The two groups of particles (A and B) living on the ring are the parts of a bi-partite system. A maximally entangled state (MES) of such a system has the form \cite{definition-MES}
\begin{equation}\label{Eq:MES}
 |MES\rangle = \sum^d_{q=1} e^{i\theta_q}|q\rangle_A |q\rangle_B,
\end{equation}
with arbitrary phases $\theta_q$, where $\{|q\rangle_D\}$ is an orthonormal basis for species $D$. Let us consider the Fock states $|q\rangle_A |q\rangle_B = |n_{1},..., n_{L}\rangle_A|n_{1},..., n_{L}\rangle_B$ where $n_j$ is the number of particles on site $j$. This state is an eigenstate of the interaction Hamiltonian with eigenvalue
\begin{equation}
 E_{int}(n_{1},..., n_{L}) = (U-V)\sum_{j=1}^L n_j^2. \nonumber
\end{equation}
We choose $V = U$ in order to achieve a degeneracy in the interaction energy for these states, i.e. $E_{int}(n_{1},..., n_{L})=0$ for any $\{n_{j}\}$. 

Let us analyze now what happens when the kinetic part of the Hamiltonian acts on $|MES\rangle$. Notice that in general the MES is not an eigenstate of $\hat{K} = \hat{K}_A + \hat{K}_B$, since $\hat{d}_{j}^{\dagger}\hat{d}_{j+1}|q\rangle_A |q\rangle_B$ does {\it not} belong to the subspace spanned by the states $|q\rangle_A |q\rangle_B$. Nevertheless, it is still possible to have $\hat{K}|MES\rangle=0$, that is, the MES is a zero-energy eigenstate.
Let us start by considering a more general state of the form
\begin{equation}
 |\psi\rangle = \sum_{\vec{n}} c_{\{n_{1},..., n_{L}\}}|n_{1},..., n_{L}\rangle_A|n_{1},..., n_{L}\rangle_B,  \nonumber
\end{equation}
where the sum is over all $\vec{n} = \{ n_1, \ldots , n_L \}$ such that $\sum_j n_j = N$. Assuming $\phi_A = \phi_B = \phi$, it is straightforward to show that
\begin{eqnarray}
\hat{K}|\psi\rangle &=& -C\sum_{j=1}^{L} \sum_{\vec{n}} (e^{i\phi}c_{\{..., n_{j}, n_{j+1},...\}} + e^{-i\phi}c_{\{..., n_{j}+1, n_{j+1}-1,...\}})\sqrt{n_{j+1}^{\phantom{,}}}\sqrt{n_j+1}   \nonumber\\
&\times&\left(|..., n_{j}+1, n_{j+1}-1,...\rangle_A|..., n_{j}, n_{j+1},...\rangle_B + |..., n_{j}, n_{j+1},...\rangle_A|..., n_{j}+1, n_{j+1}-1,...\rangle_B\right),  \nonumber
\end{eqnarray}
which implies that in order to have $\hat{K}|\psi\rangle = 0$ the coefficients must fulfill
\begin{eqnarray}
e^{i\tilde{\phi}}c_{\{..., n_{j}, n_{j+1},...\}} + e^{-i\tilde{\phi}}c_{\{..., n_{j}+1, n_{j+1}-1,...\}} = 0 \nonumber
\end{eqnarray}
for all $\vec{n}$, all $j=1,\ldots, L$, and a certain $\tilde{\phi}$. 
Furthermore, taking into account the periodic boundary conditions, we find
\begin{equation}\label{Eqn:phase}
 \tilde{\phi} =  m\frac{\pi}{L} - \frac{\pi}{2},  
\end{equation}
where $m$ is an integer. In conclusion, when $U=V$ and the phases $\phi_A = \phi_B = \tilde{\phi}$ in the Hamiltonian fulfill (\ref{Eqn:phase}), there exists a zero-energy eigenstate of the form
\begin{equation} \nonumber
 |\psi_m\rangle = \frac{1}{\sqrt{d}} \sum_{\vec{n}} e^{i\frac{2\pi m}{L}p_{\{n_{1},..., n_{L}\}}}|n_{1},..., n_{L}\rangle_A|n_{1},..., n_{L}\rangle_B,
\end{equation}
where
\begin{equation} \nonumber
 p_{\{..., n_{j}+1, n_{j+1}-1,...\}} = p_{\{..., n_{j}, n_{j+1},...\}} + 1.
\end{equation}
Notice that not only $| \psi_m \rangle$ is a {\it zero-energy eigenstate of  both the kinetic and interacting parts} of the Hamiltonian \cite{Shemesh}, but it is also a {\it maximally entangled state} of the form (\ref{Eq:MES}). This state is quite exceptional since strong fluctuations in time for $C(t)$ would have no effect over it. 

\begin{figure}
 \begin{center}
\includegraphics[width=9.0cm]{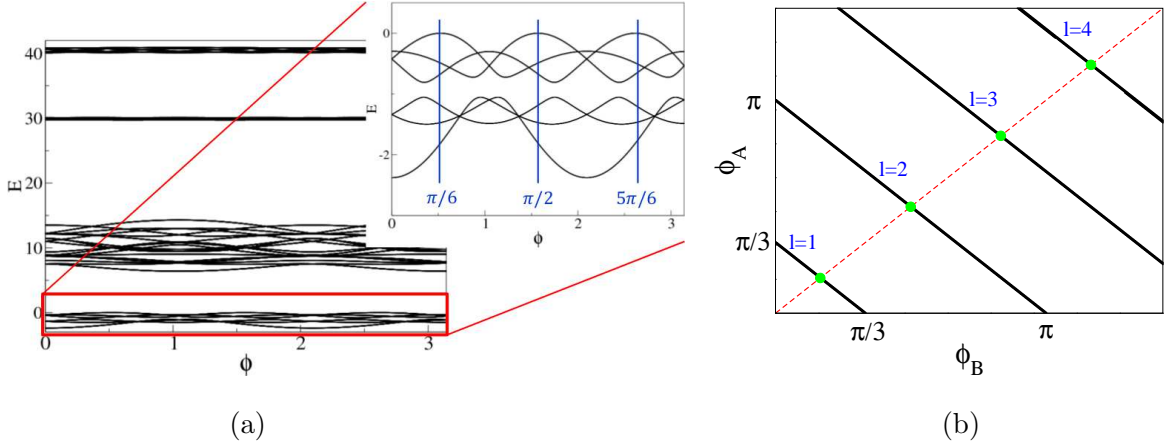}~~
\includegraphics[width=6.0cm]{MaximallyEntanglement-fase.eps}\\
~~~~(a)~~~~~~~~~~~~~~~~~~~~~~~~~~~~~~~~~~~~~~~~~~~~~~~~~~~~~~~~~~~~~~~~~~~~~~(b)
\caption{(Color online) a) Spectrum of the system for $L=3$, $N=2$ and $C=U/10=V/10$ for a range of values of $\phi$. Energy $E$ is shown in units of $C$. The inset illustrates the energies in the subspace spanned by $|n_{1},n_{2},n_{3}\rangle_A|n_{1},n_{2},n_{3}\rangle_B$, with $E_{int}(n_{1},n_{2},n_{3})=0$.  b) Continuous lines show families of maximally entangled states with zero energy at $(\tilde{\phi}_A+\tilde{\phi}_B)/2=(2l-1)\pi/6$ ($l$ integer). Dashed line corresponds to $\phi_A=\phi_B$.}
\label{Fig:spectrum}
\end{center}
\end{figure}

In what follows we consider the special case of a three-site ring ($L=3$). As shown in Fig. \ref{Fig:spectrum}a, zero-energy eigenstates are observed for $\tilde{\phi}=(2l-1)\pi/6$ ($l$ integer), in agreement with equation (\ref{Eqn:phase}). In Fig. \ref{Fig:spectrum}b it is observed that for different phases $\phi_A$ and $\phi_B$ of the two species, there are families of zero-energy MES along the lines $(\tilde{\phi}_A+\tilde{\phi}_B)/2=(2l-1)\pi/6$.

\begin{figure}
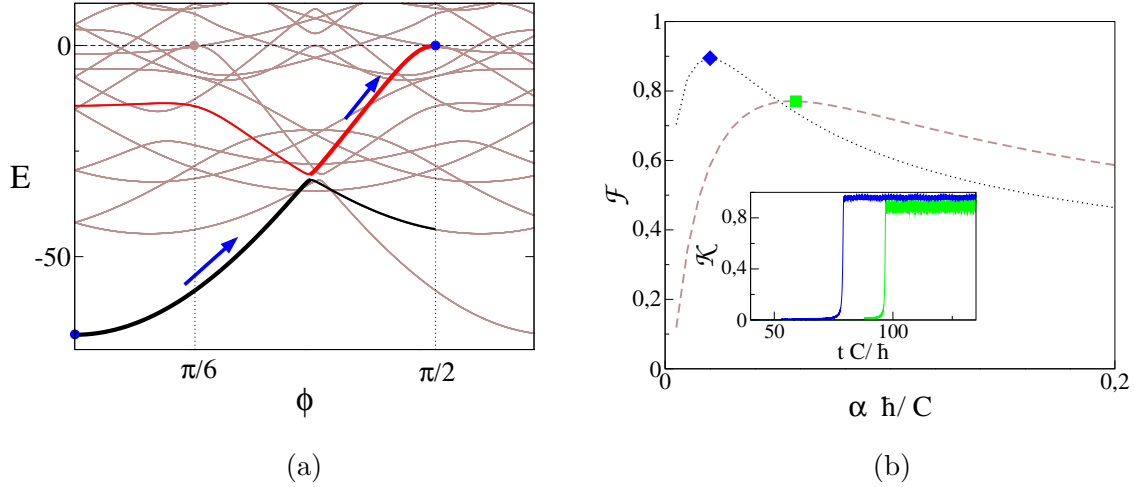

 \begin{center}
\includegraphics[width=7.0cm]{GraficoEnergia.eps}~~~~~~
\includegraphics[width=7.0cm]{Fidelity-J80-40.eps}\\
~~~~~~(a)~~~~~~~~~~~~~~~~~~~~~~~~~~~~~~~~~~~~~~~~~~~~~~~~~~~~~~~~~(b)
\caption{(Color online)  a) Energy spectrum $E$ in units of $C$ vs. $\phi$ for $L=3$, $N=2$ and $C=U$.  The two bands involved in the state preparation process are highlighted in black and in red. The blue arrows indicate the path followed to reach the zero-energy MES located at $\phi=\pi/2$. b) Fidelity $\mathcal{F}$ as a function of the velocity parameter $\alpha$. Dotted and dashed lines for $C=8U$ and $C=4U$ respectively. Diamond and square mark the maximum fidelity in each case. The time dependence of the entanglement (Schmidt number (\ref{Eq:Schmidtnumber})) during the preparation process for the values of $\alpha$ and $U$ corresponding to the diamond and square marks is shown in the inset. }
\label{Fig:LZ_spectrum}
\end{center}
\end{figure}

{\it State preparation}.---
In general, bipartite entanglement of qu$d$its is not easy to generate, and even once it is generated, it can be easily destroyed by noise, spontaneous emission, atomic collisions, etc. Nevertheless, as discussed above, in our proposed setup robust MESs exist for certain values of the Peierls phase. As evidenced on Fig. \ref{Fig:spectrum}a the MES is not the ground state of the system. Thus, the natural question that arises is how to prepare the system in such state. To this aim we propose a protocol that takes advantage of the navigation through the spectrum as the phase $\phi$, which can be varied continuously in experiments \cite{AC_force}, is changed in time. First, let us observe in Fig. \ref{Fig:LZ_spectrum}a what happens to the spectrum when choosing $C=4U$.
Note that if initially we set $\phi_i=0$, we could easily prepare the system in its ground state, since it is well separated from the first excited state. Now, let us consider a linear variation in time of the Peierls phase, $\phi(t)=\alpha t$. According to the adiabatic theorem,
for a slow variation of $\phi$ the system stays in the corresponding eigenstate (black curve in Fig. \ref{Fig:LZ_spectrum}a). In contrast, for large enough $\alpha$ there is a transition to an excited state (red curve in Fig. \ref{Fig:LZ_spectrum}a) as one passes through the avoided crossing located at $\phi_c\simeq\pi/3$. 
The probability of finding the system in the excited state after going through the avoided crossing is given approximately by the Landau-Zener probability  $P_{LZ}=e^{-\frac{2\pi\Delta^2}{\alpha \hbar}}$ \cite{Landau-Zener} where $\Delta$ is the gap at the avoided crossing. Thus, for $\alpha \gg2\pi\Delta^2/\hbar$ the transition probability is $P_{LZ} \simeq 1$ and after passing through the avoided crossing the wave function of the system will overlap strongly with the excited state. As a result, if the velocity $\alpha$ is set to zero when $\phi(t)=\pi/2$, the final state $| \psi_f \rangle$ will be very similar to the desired MES (see Fig. \ref{Fig:LZ_spectrum}a). 
A good measure of the quality of the preparation of the target state (MES) is provided by the fidelity $\mathcal{F} = |\langle MES | \psi_f \rangle|^2$. The velocity parameter $\alpha$ can be tuned to maximize $\mathcal{F}$. For instance, choosing $C=8U$ an optimal value of $\mathcal{F}\simeq 0.9$ is reached for $\alpha\simeq 0.02 C/\hbar$, as shown in Fig. \ref{Fig:LZ_spectrum}b.

To track the evolution of the degree of entanglement during the preparation process we use the so-called Schmidt number ${\cal K}_0=1/Tr_B(\rho_{B}^2)$, where $\rho_{B}$ is the reduced density matrix for subsystem $B$ \cite{Schmidt,Miguel}.
For convenience, we define the normalized Schmidt number 
\begin{equation}\label{Eq:Schmidtnumber}
{\cal K}=({\cal K}_0-1)/(d-1)\,\ ,\,\,\ 0\le {\cal K}\le 1,
\end{equation}
so that the maximum degree of entanglement is found at ${\cal K}=1$. As shown in Fig. \ref{Fig:LZ_spectrum}b an appropriate choice of parameters leads to high levels of entanglement of the order of ${\cal K}\simeq 0.95$.

\begin{figure}
 \begin{center}
\includegraphics[width=7.3cm]{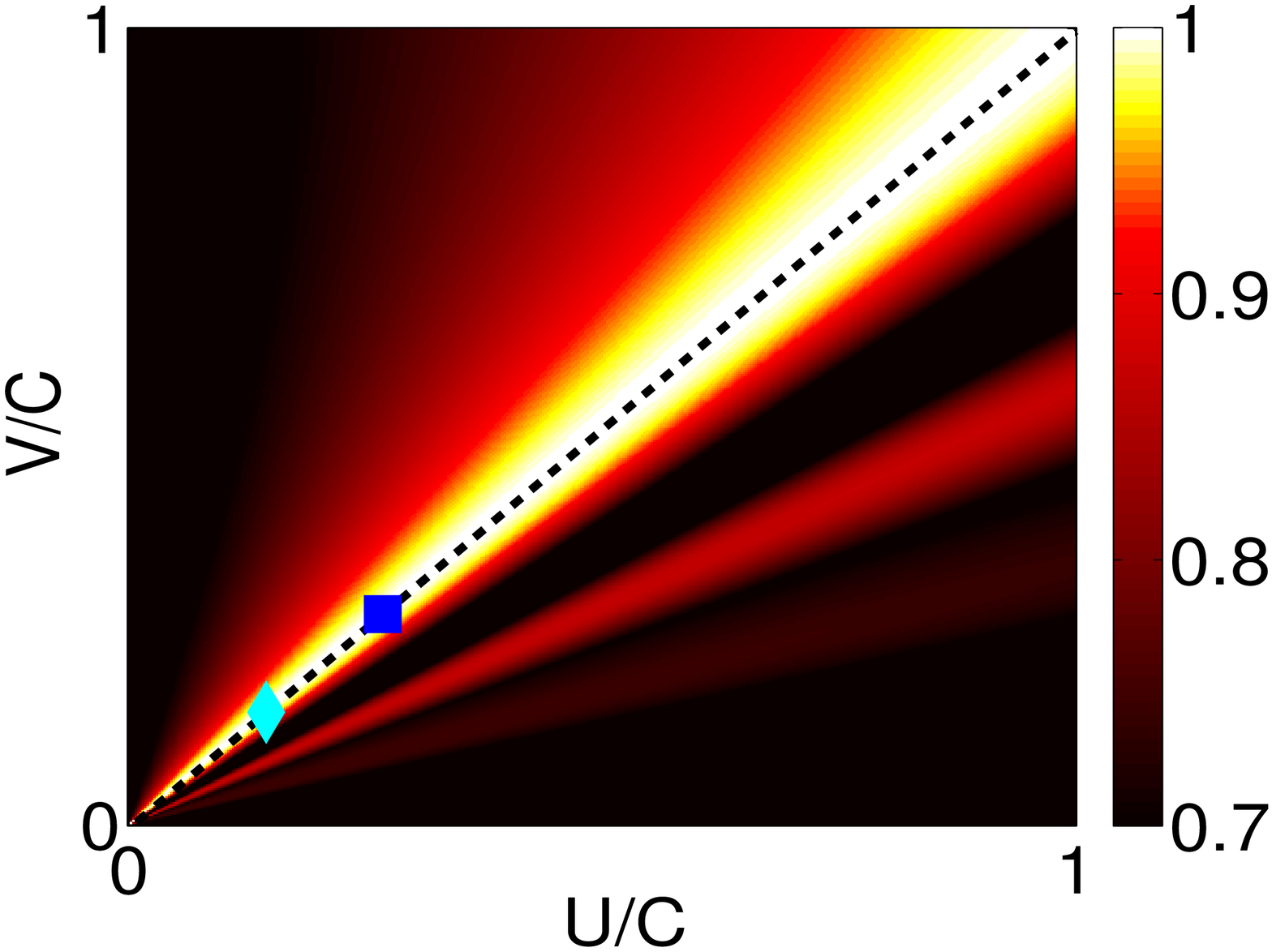}~~~~
\includegraphics[width=7.3cm]{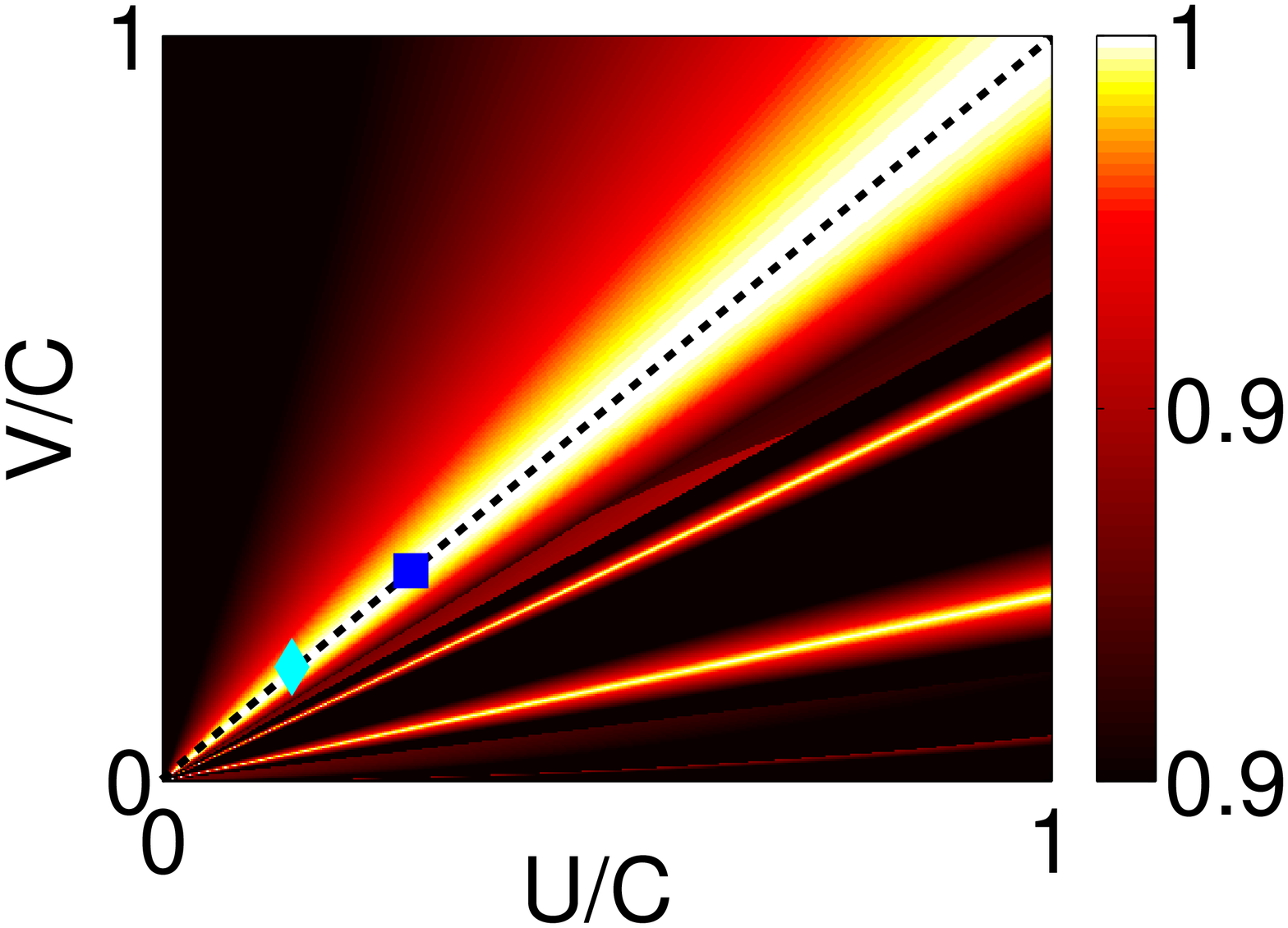}\\
(a)~~~~~~~~~~~~~~~~~~~~~~~~~~~~~~~~~~~~~~~~~~~~~~~~~~~~~~~~~~~~~~(b)
\caption{(Color online) a) Entanglement ($\mathcal{K}$) of the eigenstate with the highest entanglement as a function of the interaction parameters $U$ and $V$. b) Renormalized current ($\mathcal{J}$) of the same eigenstate as in (a). In both plots $\phi=\pi/2$ and the dashed line corresponds to $U=V$. On this line, the diamond ($\diamond$) marks the point $C=8U$, whereas the square ($\square$) indicates $C=4U$. }
\label{Fig:current}
\end{center}
\end{figure}

Further exploration of the entanglement in the $(U,V)$ parameter space for fixed $\tilde{\phi}=\pi/2$ reveals that maximally entangled eigenstates are found only when $V=U$ as displayed in Fig. \ref{Fig:current}a. Interestingly, other fringes of high entanglement are observed for $V\simeq 0.55 U$ and $V\simeq 0.25 U$.

{\it Entanglement detection}.---
The amount of entanglement is not a directly observable quantity. In this regard it has been recently shown in a similar setup that high entanglement can be linked to a low current of particles \cite{PRA_current}. The current operator for particles in the ring is $\hat{J}=\hat{J}_A\otimes \hat{\mathbf{1}}_{B}+ \hat{\mathbf{1}}_{A}\otimes \hat{J}_{B}$,
where $\hat{J}_{A,B}$ are the respective current operators for each species, defined as $\hat{J}_{D}=-\frac{iC}{\hbar L}\sum_{j=1}^{L}(e^{i\phi} \hat{d}_{j}^{\dagger}\hat{d}_{j+1}- e^{-i\phi} \hat{d}_{j+1}^{\dagger}\hat{d}_j)$ \cite{Amico}. 
In Fig. \ref{Fig:current}b we plot the renormalized current $\mathcal{J}=1-J\hbar/C$ corresponding to the eigenstate with the highest entanglement as a function of the interaction parameters $U$ and $V$. Namely, a state with low current has $\mathcal{J}\simeq 1$. Comparison with Fig. \ref{Fig:current}a shows a remarkable correlation between high entanglement and low current regions. In fact, the intensity plot for $\mathcal{J}$ displays a fringe pattern very similar to the one observed for the entanglement in Fig. \ref{Fig:current}a. Notably, at $V=U$ the current vanishes for the corresponding MES. In this way, measuring the distribution of momenta using time of flight techniques \cite{science} could provide insight into the degree of entanglement in the system.


{\it Conclusions}.---
We have developed a framework to study the generation of entanglement between two species of cold atoms living on an optical ring lattice. We have shown that synthetic magnetic fields can be exploited to create a pair of entangled qu$d$its as inter-species and intra-species interactions are set to the same value. Remarkably, for an evenly spaced discrete set of values of the Peierls phase, maximally entangled states are found. These states exhibit vanishing current and are zero-energy eigenstates of both the kinetic and interacting parts of the Hamiltonian, making them quite exceptional and robust against noise in the tunneling amplitude $C$.
Moreover, we have proposed a protocol to reach the MES based on an appropriate manipulation of the Peierls phase. Finally, an indirect detection method of the MES is proposed by measuring the current of the particles.

It is worth mentioning that recent studies in Jaynes-Cummings ring lattices have implemented synthetic gauge fields within a Bose-Hubbard framework \cite{NJP_Nunn}. Furthermore, a theoretical proposal has shown that similar physics could be observed by considering two coupled rings \cite{Amico_2}.

S.R. is supported by the FONDECYT project no 11110537. L.M.M. and S.R. acknowledge financial support from the FONDECYT project no 1110671 and M.O. is supported by the FONDECYT project no 1100039. D.S. is supported by the French project no ANR-09-BLAN-0098-01.


\begin{thebibliography}{1000} 
\bibitem{four} C. H. Bennett, G. Brassard, C. Crepeau, R. Jozsa, A. Peres, W. K. Wootters, Phys. Rev. Lett. \textbf{70,} 1895 (1993);
A. K. Ekert, Phys. Rev. Lett, \textbf{67}, 661 (1991); D. Deutsch, A. K. Ekert, R. Jozsa, C. Macchiavello, S. Popescu, A. Sanpera,
Phys. Rev. Lett, \textbf{77,} 2818 (1996).
\bibitem{interference} V. Giovannetti, S. Lloyd, and L. Macone, Phys. Rev. Lett. {\bf 96,} 010401 (2006); A. D. Cronin, J. Schmiedmayer, and D. E. Pritchard, Rev. Mod. Phys. {\bf 81,} 1051 (2009); S. Size {\it et al.}, J. Phys. B  {\bf 38,} S449 (2005).
\bibitem{PRL_Zeilinger} D. Kaszlikowski, P. Gnaci\'nski, M. Zukowski, W. Miklaszewski, and A. Zeilinger, Phys. Rev. Lett. \textbf{85,} 4418 (2000).
\bibitem{measure} R. Horodecki, P. Horodecki, M. Horodecki, and K.
Horodecki, Rev. Mod. Phys. {\bf 81}, 865 (2009).
\bibitem{three} S. F. Huelga, C. Macchiavelo, T. Pellizzari, A. K. Ekert, M. B. Plenio, J. I. Cirac, Phys. Rev. Lett, \textbf{79}, 3865 (1997).
\bibitem{Ober} O. Morsch and M. Oberthaler, Rev. Mod. Phys. {\bf 78}, 179 (2006).
\bibitem{FR} S. Inouye, et al. Nature {\bf 392}, 151 (1998). 
\bibitem{Ober0} C. Gross {\it et al}, Nature {\bf 464}, 1165 (2010). 
\bibitem{one} J. D. Weinstein, K. Beloy, A. Derevianko, Phys. Rev. A, \textbf{81}, 030302 (2010).
\bibitem{NJP_Jak} D. Jaksch and P. Zoller, New J. Phys., \textbf{5,} 56 (2003).
\bibitem{RMP_gauge} J. Dalibard, F. Gerbier, G. Juzeli\-unas and P. \"Ohberg, Rev. Mod. Phys. {\bf 83,} 1523 (2011).
\bibitem{PRL_2011_Bloch} M. Aidelsburger, {\it et al.} Phys. Rev. Lett. \textbf{107,} 255301 (2011).
\bibitem{NAT_Lin} Y.-J. Lin {\it et al.}, Nature \textbf{462,} 628 (2009).
\bibitem{PRL_Gold_1} N. Goldman {\it et al.}, Phys. Rev. Lett., \textbf{103,} 035301 (2009).
\bibitem{PRL_Gold_2} N. Goldman {\it et al.}, Phys. Rev. Lett., \textbf{105,} 255302 (2010).
\bibitem{PRL_Spiel} K. Jim\'enez-Garc\'ia {\it et al.}, Phys. Rev. Lett., \textbf{108,} 225303 (2012).
\bibitem{AC_force} J. Struck {\it et al.}, Phys. Rev. Lett., \textbf{108,} 225304 (2012).
\bibitem{PRL_2013-Ketterle} H. Miyake, {\it et al.}, Phys. Rev. Lett. \textbf{111}. 185302 (2013).
\bibitem{PRL_2013_Bloch} M. Aidelsburger, {\it et al.} Phys. Rev. Lett. \textbf{111}, 185301 (2013).
\bibitem{Ober2} Q. Y. He {\it et al. } Phys. Rev. Lett. {\bf 106}, 120405 (2011); N. Bar-Gill, C. Gross, I. Mazets, M. Oberthaler, G. Kurizki, Phys. Rev. Lett. {\bf 106}, 120404 (2011).
\bibitem{Fesch} G. Thalhammer, {\it et al.}, Phys. Rev. Lett. {\bf 100}, 210402 (2008).
\bibitem{definition-MES} N. J. Cerf, J. Mod. Opt. {\bf  47,} 187 (2000); T. Durt, D. Kaszlikowski, M. Zukowski, Phys. Rev. A {\bf 64,} 024101 (2001).
\bibitem{Shemesh} D. Shemesh, Linear Algebra and its Applications {\bf 62,} 11 (1984).
\bibitem{Landau-Zener} L. D. Landau, Physik. Z. Sowjet. \textbf{2,} 46 (1932); C. Zener, Proc. R. Soc. A \textbf{137,} 696 (1932).
\bibitem{Schmidt} R. Grobe, K. Rzazewski, J. H. Eberly, J. Phys. B {\bf 27}, L503 (1994).
\bibitem{Miguel} M. Orszag, Quantum optics, (Springer, 2008).
\bibitem{PRA_current} L. Morales-Molina, S. A. Reyes and M. Orszag, Phys. Rev. A, \textbf{86,} 033629 (2012).
\bibitem{Amico} L. Amico, A. Osterloh, and F. Cataliotti, Phys. Rev. Lett. {\bf 95}, 063201 (2005).
\bibitem{science} T. Salger, {\it et al.}, Science {\bf 326}, 1241 (2009).
\bibitem{NJP_Nunn} A. Nunnenkamp, Jens Koch and S.M. Girvin, New J. Phys., \textbf{13,} 095008 (2003).
\bibitem{Amico_2} L. Amico, D. Aghamalyan, H. Crepaz, F. Auksztol, R. Dumke and L.-C. Kwek, arXiv:1304.4615. 













\end{thebibliography}
\end{document}